\documentclass[structabstract]{aa}  
\usepackage{graphicx}
\usepackage{txfonts}
\usepackage{amssymb}
\begin{document}
   \title{On double-degenerate type Ia supernova progenitors as supersoft X-ray sources}
   \subtitle{A population synthesis analysis using SeBa}
   \titlerunning{On DD type Ia SN progenitors as SSSs}

   \author{M.T.B. Nielsen
          \inst{1,3}
          \and
	  G. Nelemans\inst{1,2}
	  \and
	  R. Voss\inst{1}
	  \and
	  S. Toonen\inst{1}
          }

   \institute{Department of Astrophysics/IMAPP, Radboud University Nijmegen,
              P.O. box 9010, 6500 GL Nijmegen, the Netherlands
	 \and
	      Institute for Astronomy, KU Leuven, Celestijnenlaan 200D, 3001 Leuven, Belgium 
         \and
	      Max-Planck Institut für Astrophysik, Karl-Schwarzschild-Str.1, Postfach 1317, D-85741 Garching, Germany\\
              \email{mede@mpa-garching.mpg.de}}

   \date{Received -; accepted -}

 
  \abstract
   {The nature of the progenitors of type Ia supernova progenitors remains unclear. While it is usually agreed that single-degenerate progenitor systems would be luminous supersoft X-ray sources, it was recently suggested that double-degenerate progenitors might also go through a supersoft X-ray phase.}
   {We aim to examine the possibility of double-degenerate progenitor systems being supersoft X-ray systems, and place stringent upper limits on the maximally possible durations of any supersoft X-ray source phases and expected number of these systems in a galactic population.}
   {We employ the binary population synthesis code SeBa to examine the mass-transfer characteristics of a possible supersoft X-ray phase of double-degenerate type Ia supernova progenitor systems for 1) the standard SeBa assumptions, and 2) an optimistic best-case scenario. The latter case establishes firm upper limits on the possible population of supersoft source double-degenerate type Ia supernova progenitor systems.}
   {Our results indicate that unlike what is expected for single-degenerate progenitor systems, the vast majority of the material accreted by either pure wind mass transfer or a combination of wind and RLOF mass transfer is helium rather than hydrogen. Even with extremely optimistic assumptions concerning the mass-transfer and retention efficiencies, the average mass accreted by systems that eventually become double-degenerate type Ia supernovae is small. Consequently, the lengths of time that these systems may be supersoft X-ray sources are short, even under optimal conditions, and the expected number of such systems in a galactic population is negligible.}
   {The population of double-degenerate type Ia supernova progenitors that are supersoft X-ray sources is at least an order of magnitude smaller than the population of single-degenerate progenitors expected to be supersoft X-ray sources, and the supersoft X-ray behaviour of double-degenerate systems typically ceases long before the supernova explosions.}


   \maketitle

%

\section{Introduction} \label{sect:Introduction}
Type Ia supernovae (SNe) are of critical importance to cosmological distance measurements and galactic evolution. Despite decades of intense research the nature of the progenitors giving rise to these explosions remains unclear (e.g. Maoz \& Mannucci \cite{Maoz.Mannucci.2012}). From observational evidence, it is inferred that the exploding objects are carbon-oxygen white dwarfs (WDs) with masses close to the Chandrasekhar mass (M$_{\mathrm{Ch}} \sim 1.38 \mathrm{M}_{\odot}$) that undergo thermonuclear runaway as carbon and oxygen is processed to radioactive iron-group elements. However, most single carbon-oxygen WDs are born at masses much smaller than M$_\mathrm{Ch}$, typically $\sim$ 0.6 M$_{\odot}$. Consequently, the fundamental problem surrounding type Ia SN progenitors revolves around how  newly-formed, initially sub-M$_\mathrm{Ch}$ WD can grow sufficiently in mass to eventually explode as type Ia SNe. It is commonly agreed that the progenitors are binary systems where the WD that eventually explodes accretes mass from a companion. Two progenitor scenarios (or 'channels') are usually considered: in the single-degenerate (SD) scenario, a single WD accretes hydrogen-rich material from a non-degenerate companion (usually a giant, although main sequence or helium-stars are also sometimes considered) and processes the accreted material to carbon and oxygen, eventually reaching the required mass where it explodes (Whelan \& Iben \cite{Whelan.Iben.1973}). In the double-degenerate (DD) scenario, a binary system consisting of two sub-M$_\mathrm{Ch}$ WDs spiral in via emission of gravitational radiation and eventually merge, forming a single carbon-oxygen WD with a combined mass at or above the required mass (Iben \& Tutukov \cite{Iben.Tutukov.1984}, Webbink \cite{Webbink.1984}). From the observational data currently available, it is not possible to clearly discriminate which scenario is the correct one, or whether both scenarios contribute to the SN Ia rate. Beyond the two main scenarios, there are a number of alternative scenarios considered by various groups, e.g. the 'core degenerate' scenario (Kashi \& Soker \cite{Kashi.Soker.2011}).

As shown by van den Heuvel et al. (\cite{van.den.Heuvel.et.al.1992}), the accretion and thermonuclear processing of H-rich material on the WD in the SD scenario is expected to emit copious amounts of supersoft X-rays ($L_{\mathrm{bol}} \sim 10^{38} \mathrm{erg/s}$ at black-body spectral fits corresponding to $T_{\mathrm{BB}}=30-150$ eV), provided the material is accreted at high enough rates (Nomoto \cite{Nomoto.1982}). This potentially makes nearby progenitor systems observable to current X-ray instruments like \textit{Chandra} and \textit{XMM-Newton}. An archival search for \textit{Chandra} pre-explosion observations at the positions of nearby type Ia SNe is being undertaken (see Voss et al. \cite{Voss.Nelemans.2008}, Nielsen et al. \cite{Nielsen.et.al.2012}, Nielsen et al. \cite{Nielsen.et.al.2013a}), but so far, no unambiguous direct detections of supersoft X-ray sources (SSSs) at the positions of type Ia SNe have been made. Additionally, when compared to the population that should be expected if the SD scenario is responsible for the observed SN Ia rate, the observed population of SSSs in nearby galaxies falls short by at least one, and quite likely two orders of magnitude (Di Stefano \cite{Di.Stefano.2010.a}). Likewise, the integrated soft X-ray luminosity measured from elliptical galaxies falls similarly short (Gilfanov \& Bogd{\'a}n \cite{Gilfanov.Bogdan.2010}), assuming SSS SD systems are the progenitors of type Ia SNe. Taken at face value, these points should be considered serious problems for the SD scenario (however, see the Discussion section for alternative explanations of the absence of SSSs).

To complicate matters further, it has been suggested that even if the DD scenario is the dominant one in terms of supplying progenitors of type Ia SNe, a large population of SSSs would still be expected to exist (Di Stefano \cite{Di.Stefano.2010.b}). The motivation for this is that most of the binary systems that eventually become DD progenitors of type Ia SNe should pass through a stage where they consist of a WD and a non-degenerate companion, before the latter becomes a WD. This configuration mimics the late stages of a SD system where it could be a SSS. If DD progenitor systems are also SSSs for a significant amount of time ($\sim$ Myr), there could be an observationally significant number of such systems at any one time in a galactic population like the Milky Way. If correct, this would mean that the absence of a large population of SSS could potentially be a problem for both progenitor scenarios, not just for the SD scenario.

If we wish to understand the nature of the progenitors of type Ia SNe, we must obtain a better understanding of the observational characteristics of the progenitor scenarios. We need to settle whether either of the scenarios give rise to supersoft X-ray emission. In the present study, we examine the hypothesis that DD progenitors are SSS, using the SeBa binary evolution code (Portegies Zwart et al. \cite{Portegies.Zwart.1996}, Nelemans et al. \cite{Nelemans.et.al.2001}, Toonen et al. \cite{Toonen.et.al.2012}). In Section \ref{sect:Theory} we review the theory behind SSSs and the proposed SSS nature of DD type Ia SN progenitor systems. In Section \ref{sect:Method} we explain the details of our method. Section \ref{sect:Results} lists our results, and Section \ref{sect:Discussion} discusses the implications of these results. Section \ref{sect:Conclusion} concludes.

A word on terminology: we use 'mass transfer' to denote the transfer of material from donor to accretor, regardless of whether some of that material is subsequently lost from the accretor. By 'retention efficiency' we mean the fraction of the transferred material that remains on the accretor. By 'accretion' we refer to transferred material that remains on the accretor. So, a mass-transfer rate of $10^{-7}$ M$_{\odot}$/yr that is retained at 25\% retention efficiency results in an accretion rate of $2.5 \cdot 10^{-8}$ M$_{\odot}$/yr, for example. Note that retention efficiency and accretion efficiency are not the same; the accretion efficiency is the ratio of the total amount of matter lost from the donor that remains on the accretor, while retention efficiency is the ratio of the transferred material that remains on the accretor.

%

\section{Theory} \label{sect:Theory}
For an initially sub-M$_{\mathrm{Ch}}$ WD to grow in mass and eventually become a type Ia SN, material from the donor star needs to be transferred and retained on the WD. While a wide range of mass-transfer rates are possible, the retention of transferred material depends on the mass of the accretor and the mass-transfer rate; for WDs above 0.6 M$_{\odot}$, the transfer of hydrogen-rich material can only take place in a stable manner in a narrow interval of mass-transfer rates, between $1.7 \cdot 10^{-7}$ M$_{\odot}$/yr and $4.1 \cdot 10^{-7}$ M$_{\odot}$/yr (Nomoto \cite{Nomoto.1982}). Outside of this interval, the transferred material is unlikely to be retained on the WD; for mass transfer rates above the steady-burning rate, the material is transferred onto the WD faster than it can be processed, and the WD consequently swells up, likely stopping or severely hampering the mass transfer process, see also Nomoto et al. (\cite{Nomoto.et.al.1979}). For mass-transfer rates below the steady-burning rate, the material burns unstably (Fujimoto \& Sugimoto \cite{Fujimoto.Sugimoto.1979}, \cite{Fujimoto.Sugimoto.1982}), i.e. in nova eruptions, causing the WD to lose most of the accreted mass, plus possibly some additional mass from the WD itself.

A similar constraint governs the mass transfer of helium-rich material, i.e. very high mass-transfer rates cause the accretor to swell up, somewhat lower mass-transfer rates allow steady burning, while low mass-transfer rates result in helium-novae. The question of helium steady burning was examined by Hachisu et al. (\cite{Hachisu.et.al.1999}), Kato \& Hachisu (\cite{Kato.Hachisu.1999}), and Iben \& Tutukov (\cite{Iben.Tutukov.1996}) (see also review by Bours et al. \cite{Bours.et.al.2013}). Because of the higher temperatures and densities required for helium burning, higher mass-transfer rates are required for helium to burn steadily, as compared to hydrogen mass transfer. The exact value of the steady burning rate is somewhat disputed, but for a 1 M$_{\odot}$ WD, the interval of steady burning mass-transfer rates that agrees with all of three studies mentioned above is between $2.5 \cdot 10^{-6}$ M$_{\odot}$/yr and $4.0 \cdot 10^{-6}$ M$_{\odot}$/yr (see Fig.2 in Bours et al. \cite{Bours.et.al.2013}).

To get the initially sub-M$_{\mathrm{Ch}}$ to the mass needed for a type Ia SN explosion in the SD scenario, an extended period of steady mass transfer is required after the formation of the WD. Since carbon-oxygen WDs are not expected to form at masses larger than $1-1.2$ M$_{\odot}$, the steady mass transfer and processing of material must last on the order of a few million years or longer. The mechanism of mass transfer can be anything that is capable of supplying a transfer of matter at the steady-burning rate; usually, it is assumed to happen either through a wind or by Roche-lobe overflow (RLOF).

In the case of DD progenitors, a binary system evolves to consist first of a single WD and a non-degenerate companion, and later two WDs that eventually merge to form a single WD with a mass sufficient to explode as a type Ia SN (however, see our comment concerning 'double-CE' systems in Section \ref{sect:Method} below). At some intermediate point during its evolution, before the merger happens, such proto-DD systems will consist of a WD and a non-WD companion star, and hence may be considered conceptually similar to a SD system. Since we expect SD type Ia SN progenitor systems to be SSSs as a result of the thermonuclear processing of the accreted material, we may also consider the prospect that such 'SD-like', proto-DD type Ia SN progenitors could display similar behaviour in this phase of their evolution, if they accrete material from their companions at sufficiently high rates. In Di Stefano (\cite{Di.Stefano.2010.b}) it was suggested that a significant fraction of these systems should be expected to accrete H-rich material from their companions at a rate large enough to sustain steady burning, corresponding to a population of 'thousands' in a spiral galaxy like the Milky Way. They would therefore also emit supersoft X-rays, similarly to a SSS SD type Ia SN progenitor system, for an extended period of time ($\sim$ Myr). The mechanism behind this mass transfer is wind mass transfer, and Di Stefano (\cite{Di.Stefano.2010.b}) assumed a wind accretion efficiency of 25\%, i.e. one-fourth of the material lost from the non-degenerate companion is accreted onto the WD.

Two key requirements need to be met for the SD-like proto-DD type Ia SN progenitor systems to constitute a significant population of observable SSSs. Firstly, the mass-transfer rate needs to be high enough for the transferred material to be burned steadily on the surface of the WD, thereby giving rise to supersoft X-ray emission. If this requirement is not met, the sources may still accrete material (albeit at much smaller retention efficiency, as described in Nomoto \cite{Nomoto.1982}), but they will presumably not emit much in terms of supersoft X-rays. Secondly, the accretion of material at the steady-burning rate needs to take place over a long enough period of time, so that at any one time there will be a significant population of these sources present for us to observe.

Before the merger and SN can take place the second WD needs to form, after which the decay of the orbit will lead to the merger. Due to the time needed for this process (during which the system no longer is SD-like and not expected to emit supersoft X-rays), it will not be possible to directly associate a given SN with a previously observed SSS if DD systems are the dominant progenitor channel for type Ia SNe.

%

\section{Method} \label{sect:Method}
We employed the binary population synthesis code SeBa (Portegies Zwart et al. \cite{Portegies.Zwart.1996}, Nelemans et al. \cite{Nelemans.et.al.2001}, Toonen et al. \cite{Toonen.et.al.2012}) to simulate the evolution of a large number of binaries. The evolution is followed for a Hubble time starting from the zero-age main-sequence. At every timestep, stellar winds, mass transfer, common envelopes (CEs), angular momentum loss, and gravitational waves are taken into account with appropriate recipes. We assume solar metalicities, and the initial primary masses are distributed according to the Kroupa initial mass function (Kroupa et al. \cite{Kroupa.et.al.1993}) between 0.95-10 M$_{\odot}$ and the initial mass ratio distribution is flat. The distribution of orbital separations is flat in log-space (Abt \cite{Abt.1983}) out to $10^6$ R$_{\odot}$, and the eccentricities are distributed thermally between 0 and 1 (Heggie \cite{Heggie.1975}). Due to uncertainties in the physics of CEs (for an overview, see Ivanova et al. \cite{Ivanova.et.al.2013}), several prescriptions for the CE-phase exist that are based on the energy budget (the $\alpha$-prescription, see Tutukov \& Yungelson \cite{Tutukov.Yungelson.1979}, Webbink \cite{Webbink.1984}) or on the angular momentum balance (the $\gamma$-prescription, see Nelemans et al. \cite{Nelemans.et.al.2000}). In SeBa, the $\alpha$-formalism is used in all cases where the binary contains a compact object, or when a CE is triggered by a tidal instability. For all other CE-events, the $\gamma$-formalism is used. The results given below (in the main text, not in the Appendix) use these assumptions. For both the standard and optimistic cases (see below) we assume $\gamma = 1.75$ (Nelemans et al. \cite{Nelemans.et.al.2001}) and $\alpha\lambda = 2$ (Nelemans et al. \cite{Nelemans.et.al.2001}), where $\lambda$ is the envelope-structure parameter (de Kool et al. \cite{de.Kool.et.al.1987}). For completeness, in the Appendix we also include results for a model that applies the $\alpha$-formalism to all CE-events.

We note that for the $\alpha$-CE formulation, there are systems that develop directly into DD systems from the giant phase, thereby avoiding the SD-like phase; for the standard $\gamma$-CE formulation used here, this does not happen. See Toonen et al. \cite{Toonen.et.al.2012}) for details.

We ran a SeBa simulation for a total of 500,000 binary systems. From the resulting outputs we conducted analyses for two separate cases: a 'standard' and an 'optimistic' case. In the former, we simply took the SeBa outputs at face value. In the latter case, we manually imposed optimistic conditions concerning transfer and retention of material (see below). The motivation for the second analysis was to calculate a solid upper limit for the populations of DD progenitor systems that could possibly be SSSs, and specifically to compare with the results of Di Stefano (\cite{Di.Stefano.2010.b}), whose study used more generous assumptions concerning the efficiency of wind mass transfer than assumed by SeBa. The reason it is possible to manually impose a different retention efficiency from the SeBa outputs in the optimistic case is that SeBa explicitly outputs the mass loss from both binary components in each calculation step. By finding the difference in donor mass in each step in which the donor is not transferring mass stably we can find the amount of material lost in a wind. We then assume a given retention efficiency to find the fraction of this material that ends up being accreted onto the donor. As long as the masses accreted in this way are small compared to the mass of the accretor - which they always are - this approach does not significantly change the general physical and evolutionary behaviour of the binaries, which means that the subsequent SeBa steps are still correct.

In both cases, we examined all systems consisting of a single WD and a non-degenerate companion that would later merge to a final mass above the Chandrasekhar mass, i.e. systems that could be said to be SD-like before becoming DD type Ia SNe. We calculated the accreted masses of both hydrogen- and helium-rich material.

In the standard case, the masses accreted from wind and RLOF is directly given by the code. For wind mass transfer, SeBa only considers Bondi-Hoyle-Lyttleton wind accretion, the accretion efficiency of which is quite small (typically, $<$ 1\%, see Edgar \cite{Edgar.2004}). For both wind and RLOF mass transfer, SeBa follows the steady-burning constraints of Nomoto (\cite{Nomoto.1982}), i.e. material transferred at rates different from the steady-burning rate is unlikely to be appreciably retained on the WD. We refer to Bours et al. (\cite{Bours.et.al.2013}) for further details on the assumptions concerning wind and stable mass transfer in SeBa.

For the optimistic case, we relaxed the assumptions concerning both wind and RLOF mass transfer to enable comparison with Di Stefano (\cite{Di.Stefano.2010.b}) and establish upper limits on the possible lifetime and number of SSS proto-DD type Ia SN progenitor systems. For wind transfer, we counted the total mass transferred as the mass lost from the donor star while the the accretor is a WD and the binary components are detached (i.e. not in a CE or inspiraling phase). Only a fraction of this material will actually be accreted by the accretor, and the rest of it will be lost from the system. The exact amount accreted depends on the model used. Di Stefano (\cite{Di.Stefano.2010.b}) assumed an accretion efficiency of 25\% for wind mass transfer. To get strong upper limits we adopted the same wind accretion efficiency as Di Stefano, i.e. 25\%. For both wind and RLOF we assumed a retention efficiency of 100\%, i.e. all the mass that ends up on the accretor stays there. This is obviously quite an optimistic assumption, since, as mentioned above, the material needs to be accreted at a fairly narrow range of mass-transfer rates in order to facilitate full retention. However, in the context of the current study, we are content to establish an upper limit of the number of SD-like proto-DD type Ia SN progenitor systems that could conceivably be SSSs.

If we assume that all the accreted material (hydrogen or helium) is transferred at the steady-burning rate appropriate for that type of material, the average SSS lifetime $\tau_{\mathrm{accr}}$ of a DD SN progenitor in a given stellar population is given by:
\begin{eqnarray}
 \tau_{\mathrm{accr}}/\mathrm{SN} = \sum^{H,He}_{X} \frac{\Delta M_{X}}{\dot{M}_{X\mathrm{,steady}}}
\end{eqnarray}
where $\Delta M_{X}$ is the total accreted mass of material $X$, and $\dot{M}_{X\mathrm{,steady}}$ is the minimum mass-transfer rate required for steady-burning of material $X$.

The donors in SeBa can be either hydrogen- or helium-rich. As mentioned in Section \ref{sect:Theory}, for helium-rich material, the mass-transfer rate needed to sustain steady burning and avoid significant mass loss through nova eruptions is roughly an order of magnitude larger than for hydrogen-rich material. This larger steady-burning mass-transfer rate translates into a shorter SSS life-time for the same mass of material, as compared with a system transferring hydrogen. Since we want to determine an upper limit to the number of possible DD progenitors that can be SSSs at any given time, we take the minimum steady-burning rates mentioned above, i.e. $\dot{M}_{\mathrm{H,steady}}=1.7 \cdot 10^{-7}$ M$_{\odot}$/yr and $\dot{M}_{\mathrm{He,steady}}=2.5 \cdot 10^{-6}$ M$_{\odot}$/yr. For simplicity, we assume the material transferred from H-rich donors (i.e. main sequence stars, Herzsprung gap stars, first giant branch stars, core helium-burning stars, and asymptotic giant stars) to consist of a 25\% helium and 75\% hydrogen (by mass), while the material transferred from helium-rich systems (helium-stars and helium-giants) is exclusively helium.

The average number of sources $N_{\mathrm{accr}}$ is calculated by scaling the average SSS lifetime with the average occurrence rate of type Ia SNe in a galaxy:
\begin{eqnarray}
 N_{\mathrm{accr}} =  3.0 \cdot 10^{-3} \tau_{\mathrm{accr}} \left( \frac{L_{\mathrm{B}}}{10^{10} \mathrm{L}_{\mathrm{B,}\odot}} \right) \label{eq:N_accr}
\end{eqnarray}
where $L_{\mathrm{B}}$ is the B-band luminosity of the galaxy, L$_{\mathrm{B,}\odot}$ is the B-band luminosity of the Sun, and we have assumed an type I SN rate of 3 per millennium, typical of a spiral galaxy like the Milky Way. We limit ourselves to considering a population similar to the Milky Way, which means that the last term in Eq.(\ref{eq:N_accr}) is equal to 1.

%

\section{Results} \label{sect:Results}
In this section, we present the results for the standard and the optimistic cases. The former gives realistic estimates of the mass accretion, according to our best current understanding. The latter gives strong upper limits to the mass accretion which should be applicable no matter which assumptions are made concerning mass-transfer and retention efficiencies.

Of the simulated 500,000 binary systems, 2290 systems resulted in double carbon-oxygen WD mergers with a combined mass above 1.38 M$_{\odot}$ for the $\gamma$-CE prescription in the standard case (see Section \ref{sect:Method} for further information on when the $\alpha$- and $\gamma$-formalisms are used in SeBa). When we relaxed our mass-transfer and retention efficiency assumptions, 175 systems that were DD type Ia progenitors in the standard case experience enough mass transfer to bring their WDs above 1.38 M$_{\odot}$ before the systems merge. Under our optimistic assumptions, these systems would therefore become SD (instead of DD) type Ia SNe. Consequently, we removed them from our optimistic case sample, leaving us with 2115 systems for this case.

For general applicability, rather than giving our results in total numbers, we list them in terms of solar masses per SN. This enables the reader to scale the results with any particular supernova rate and progenitor life-time of their choice. The total masses can be found by multiplying the average numbers by the number of progenitor systems in each sample.

In the following, we only discuss the results for the $\gamma$-CE formulation. However, the results for the $\alpha$-CE are quite similar and yield essentially the same conclusions. The results for the $\alpha$-CE are given in the Appendix.

\subsection{All systems} \label{subsect:Results_All}
\begin{figure}[ht]
\centerline{\includegraphics[width=\linewidth]
{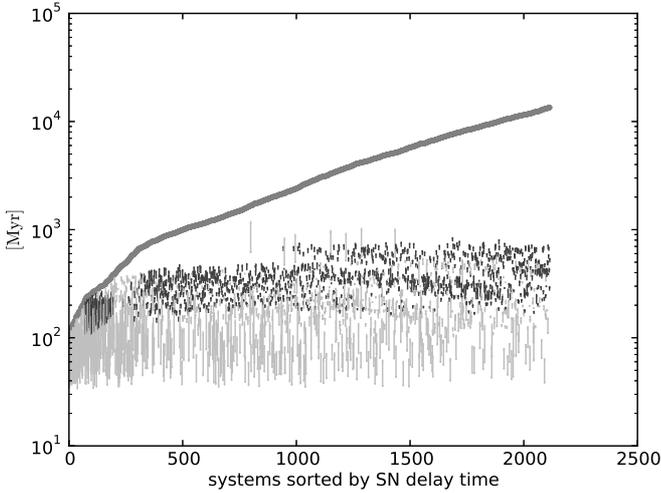}}\caption{Accretion history for all systems in the optimistic case, using the $\gamma$-CE formulation. The y-axis gives the delay time from formation of the system until the SN explosion. Each column in the plot corresponds to one system, and the systems are ordered according to delay time, with delay times increasing towards the right; the dark grey line delineates the SN explosions. Light grey vertical lines show accretion events in systems where the initially most massive star is the accretor, black is for systems where the initially least massive star is the accretor. Compare with Figures \ref{fig:accr_history_log_4_wind} \& \ref{fig:accr_history_log_4_RLOF}}
\label{fig:accr_history_log_4_all}
\end{figure}

The average masses accreted per SN in our entire sample are given in Table \ref{table:deltaM_donor_types_all}. Figure \ref{fig:accr_history_log_4_all} shows the accretion histories of these systems. Clearly, helium-accreting systems dominate in terms of the amount of mass being transferred for both the standard and optimistic assumptions.

Table \ref{table:deltaM_donor_types_all} also gives the average SSS life-time of the progenitor systems, based on the average mass accreted per SN in the total sample. This SSS life-time assumes that all material is burned at the steady-burning rate, and takes the different steady-burning rates for hydrogen and helium into account. 
For both the standard and optimistic cases, the average life-times are roughly 0.05 Myr, significantly smaller than both the expected total life-time of an average DD type Ia SN progenitor system (e.g. Maoz et al. \cite{Maoz.et.al.2010}), and the expected supersoft X-ray life-time of SD progenitors. The SSS life-times translate into a number of accreting SSS systems expected to be 'on' in Milky-Way type spiral galaxy at any time, according to Eq.(\ref{eq:N_accr}). Table \ref{table:deltaM_donor_types_all} lists this number for both of the examined cases. We also list the Poissonian errors on these numbers, with the caveat that at the end of the day, the accreted masses, and thereby the calculated $N_{\mathrm{accr}}$ depend on the assumptions used in SeBa. For a detailed discussion of these assumptions we refer to Toonen et al. (\cite{Toonen.et.al.2012})

\begin{table}
\caption{Mass accreted by the first-formed WDs in all DD progenitor systems (2290 for the standard case; 2115 for the optimistic case), split by donor type. The bottom rows give the SSS life-time of the average $\Delta M$, if all material is accreted at the component-specific (for H and He, respectively) steady-burning mass-transfer rates, and the resulting expected number of accreting systems (with Poisson errors). Both columns use the $\gamma$-CE prescription.}
 \centering
  \begin{tabular}{c c c}
  \hline
					& standard		& upper limit	\\
  					& SeBa case:		& case:	\\
  donor stellar				& $\Delta M$/SN		& $\Delta M$/SN		\\
  type					& [M$_{\odot}$]		& [M$_{\odot}$]		\\
  \hline
  \hline
   main sequence star			& 0.0			& 5.98 $\cdot10^{-5}$	\\
   Herzsprung gap star			& 0.0			& 2.61 $\cdot10^{-5}$	\\
   first giant branch star		& 7.14 $\cdot10^{-5}$	& 2.87 $\cdot10^{-5}$	\\
   core He-burning star			& 0.0			& 1.81 $\cdot10^{-4}$	\\
   asymptotic giant branch star	& 0.0			& 3.91 $\cdot10^{-5}$	\\
   He-star				& 4.98  $\cdot10^{-5}$	& 2.31 $\cdot10^{-3}$	\\
   He-giant star			& 8.54 $\cdot10^{-2}$ 	& 1.17 $\cdot10^{-1}$	\\
   Total, all types			& 8.55 $\cdot10^{-2}$	& 1.19 $\cdot10^{-1}$	\\
  \hline
  \hline
   $\tau_{\mathrm{accr}}$ [yr]		& 3.5 $\cdot10^{4}$	& 4.9 $\cdot10^{4}$	\\
   $N_{\mathrm{accr}}$ ($10^{10}$ L$_{\mathrm{B,}\odot}$ galaxy)	& 1.0 $\cdot10^{2} \pm$ 10	& 1.5 $\cdot10^{2} \pm$ 12	\\
  \hline
\end{tabular} \label{table:deltaM_donor_types_all}
\end{table}

Figure \ref{fig:accr_history_log_4_all} shows the accretion history for all the systems in our sample for the optimistic case. Supersoft behaviour can only take place during periods where the systems are accreting. As can be seen, systems where the initially most massive star evolves into a WD first generally finish transferring mass earlier than is the case for the systems where the initially least massive star evolves into a WD first. This is to be expected, since in the latter type of system, the initially most massive star becomes a long-lived helium-star, and so the initially least massive star needs time to evolve to a WD before mass transfer can start. This type of 'evolution-reversed' systems will therefore be slow to form, and will not start transferring mass until somewhat later than systems where the most massive star becomes a WD first (see Toonen et al. \cite{Toonen.et.al.2012} for more details on this type of evolution).

Table \ref{table:time_from_last_accr_to_SN_-_all} gives the time from the last mass transfer event until the SN explosion for all the systems in our sample for the optimistic assumptions. For the vast majority (99.999\%) of the systems, mass transfer ceases a Myr or more before the SN explosion. So, even under the optimistic assumption that all of the involved systems are transferring mass at exactly the right steady-burning rate to emit supersoft X-rays, it would not be possible to observationally associate any of these systems with SN explosions, as they would have ceased to be SSSs long (millions to billions of years) before the SNe take place. This is expected, as the second white dwarf needs time to form before the system can merge and explode.

\begin{table}
\caption{Time from last mass transfer event to SN explosion, all accreting systems, optimistic case. Compare with tables \ref{table:time_from_last_accr_to_SN_-_wind} and \ref{table:time_from_last_accr_to_SN_-_RLOF}.}
 \centering
  \begin{tabular}{c c c}
  \hline
  time since			& no. of	& fraction of		\\
  last accr			& systems	& total			\\
  \hline
   t $<$ 1 Myr			& 2		& 9.46 $\cdot10^{-4}$	\\
   1 Myr $< t <$ 10 Myr		& 14		& 6.62 $\cdot10^{-3}$	\\
   10 Myr $< t <$ 100 Myr	& 117		& 5.53 $\cdot10^{-2}$	\\
   100 Myr $< t <$ 400 Myr	& 171		& 8.09 $\cdot10^{-2}$	\\
   t $>$ 400 Myr		& 1811		& 8.56 $\cdot10^{-1}$	\\
  \hline
  Total				& 2115		& 1.00			\\  
  \hline
\end{tabular} \label{table:time_from_last_accr_to_SN_-_all}
\end{table}

\subsection{Wind accreting / symbiotic systems}
Some systems never experience stable mass transfer from the donor to the accretor while they are in the SD-like configuration, and instead accrete exclusively via a wind. Such systems could be considered roughly similar to symbiotic SD type Ia supernova progenitors during this phase.

Table \ref{table:deltaM_donor_types_wind} lists the mass accreted per supernova for all systems of this type. As expected, wind mass transfer is a negligible contributor to WD growth in the standard case, so we expect no SSSs powered by this type of mass transfer for that case. With the relaxed assumptions in the optimistic case, the systems do accrete some material (and a small number become SD type Ia SNe as a result, as mentioned), although still quite small amounts compared to systems experiencing RLOF (see below). The majority of the accreted material is helium-rich. The average amount of material accreted for this type of system corresponding to an average SSS life-time of roughly 5000 years. This is completely negligible compared to both the total life-times of even the shortest-living systems that become DD type Ia SNe, and the expected SSS phase of a SD progenitor. The expected number of these SSSs active in a galactic population is therefore also tiny. The considerations concerning errors on $N_{\mathrm{accr}}$ mentioned in Subsection \ref{subsect:Results_All} are clearly applicable for Table \ref{table:deltaM_donor_types_wind} as well.

\begin{table}
\caption{Mass accreted via wind mass transfer by the first-formed WDs in DD progenitor systems in our sample (924 for the standard case; 902 for the optimistic case), split by donor type. The bottom row gives the SSS life-time of the average $\Delta M$, if all material is accreted at the component-specific (for H and He, respectively) steady-burning mass-transfer rates, and the resulting expected number of accreting systems (with Poisson errors). Both columns use the $\gamma$-CE prescription.}
 \centering
  \begin{tabular}{c c c}
  \hline
					& standard		& upper limit	\\
					& SeBa case:		& case:	\\
  donor stellar				& $\Delta M$/SN		& $\Delta M$/SN		\\
  type					& [M$_{\odot}$]		& [M$_{\odot}$]		\\
  \hline
   main sequence star			& 0.0 			& 0.0			\\
   Herzsprung gap star			& 0.0 			& 2.44 $\cdot10^{-5}$	\\
   first giant branch star		& 0.0 			& 4.23 $\cdot10^{-5}$	\\
   core He-burning star			& 0.0 			& 3.21 $\cdot10^{-6}$	\\
   asymptotic giant branch star	& 0.0 			& 2.47 $\cdot10^{-6}$	\\
   He-star				& 0.0 			& 1.03 $\cdot10^{-3}$	\\
   He-giant star			& 0.0 			& 3.50 $\cdot10^{-3}$	\\
   total, all types			& 0.0			& 4.60 $\cdot10^{-3}$	\\
  \hline
  \hline
   $\tau_{\mathrm{accr}}$ [yr]		& 0.0			& 2.16 $\cdot 10^{3}$	\\  
   $N_{\mathrm{accr}}$  ($10^{10}$ L$_{\mathrm{B,}\odot}$ galaxy)	& 0			& 6.5 $\pm$ 2.5 \\
  \hline
\end{tabular} \label{table:deltaM_donor_types_wind}
\end{table}

\begin{figure}[ht]
\centerline{\includegraphics[width=\linewidth]
{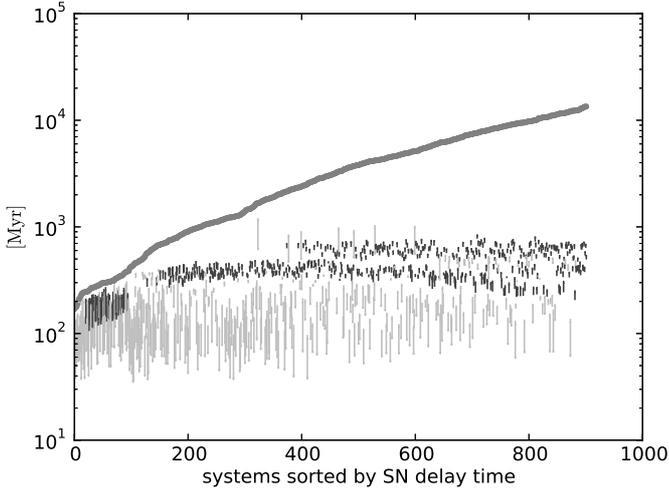}}\caption{Accretion history for detached, purely wind-accreting systems (i.e. systems that never experience stable mass transfer) in the optimistic case, using the $\gamma$-CE formulation. The y-axis gives the delay time from formation of the system until the SN explosion. Each column in the plot corresponds to one system, and the systems are ordered according to delay time, with delay times increasing towards the right; the dark grey line delineates the SN explosions. Light grey vertical lines show accretion events in systems where the initially most massive star is the accretor, black is for systems where the initially least massive star is the accretor. Compare with Figures \ref{fig:accr_history_log_4_all} \& \ref{fig:accr_history_log_4_RLOF}.}
\label{fig:accr_history_log_4_wind}
\end{figure}

Figure \ref{fig:accr_history_log_4_wind} shows the accretion history of the systems of this type, and
Table \ref{table:time_from_last_accr_to_SN_-_wind} gives the time from the last mass transfer event until the SN explosion for the optimistic assumptions. For systems transferring mass via a wind, mass transfer ceases at least 10 Myrs before the SN explosion. For $\sim$83\% of the systems, this time interval is larger than 100 Myrs. Wind accretors generally finish their accretion earlier than RLOFing systems (see next section).

\begin{table}
\caption{Time from last mass transfer event to SN explosion, detached / wind-accreting systems, optimistic case. Compare with tables \ref{table:time_from_last_accr_to_SN_-_all} and \ref{table:time_from_last_accr_to_SN_-_RLOF}.}
 \centering
  \begin{tabular}{c c c}
  \hline
  time since			& no. of	& fraction of		\\
  last accr			& systems	& total			\\
  \hline
   t $<$ 1 Myr			& 0		& 0.0			\\
   1 Myr $< t <$ 10 Myr		& 0		& 0.0			\\
   10 Myr $< t <$ 100 Myr	& 56		& 6.21 $\cdot10^{-2}$	\\
   100 Myr $< t <$ 400 Myr	& 94		& 1.04 $\cdot10^{-1}$	\\
   t $>$ 400 Myr		& 752		& 8.34 $\cdot10^{-1}$	\\
  \hline
  Total				& 902		& 1.00			\\  
  \hline
\end{tabular} \label{table:time_from_last_accr_to_SN_-_wind}
\end{table}

\subsection{RLOF-accreting systems}
Table \ref{table:deltaM_donor_types_RLOF} lists the mass accreted per supernova for systems that experience mass transfer via RLOF, at some point during their evolution. All of these systems also experience wind mass transfer at some point.

As with wind accreting systems, RLOF-transferring systems accrete predominantly helium and only negligible amounts of hydrogen. The average mass accreted by these systems is significantly larger than that accreted by purely wind-accreting systems. The expected SSS life-time of these systems is correspondingly larger, slightly less than $10^{5}$ years, however, this is still significantly less than the expected total life-time of a DD type Ia SN progenitor system, and at least an order of magnitude smaller than the SSS period expected for SD progenitor systems. The expected number of these systems in a Milky-Way type galactic population at any time is therefore still smaller than what was estimated in Di Stefano (\cite{Di.Stefano.2010.b}). See subsection \ref{subsect:Results_All} for considerations concerning errors on $N_{\mathrm{accr}}$.

\begin{figure}[ht]
\centerline{\includegraphics[width=\linewidth]{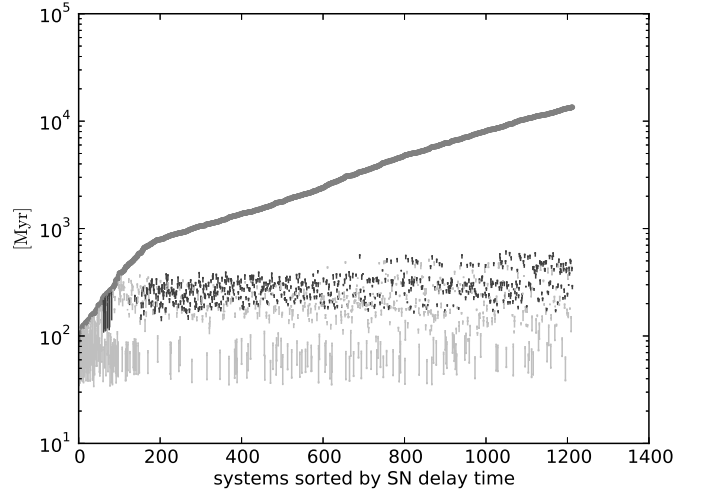}}
\caption{Accretion history for systems that experience stable mass transfer at some point during their evolution in the optimistic case, using the $\gamma$-CE formulation. Most of the systems in this category will also experience wind mass transfer. The y-axis gives the delay time from formation of the system until the SN explosion. Each column in the plot corresponds to one system, and the systems are ordered according to delay time, with delay times increasing towards the right; the dark grey line delineates the SN explosions. Light grey vertical lines show accretion events in systems where the initially most massive star is the accretor, black is for systems where the initially least massive star is the accretor. Compare with Figures \ref{fig:accr_history_log_4_all} \& \ref{fig:accr_history_log_4_wind}.}
\label{fig:accr_history_log_4_RLOF}
\end{figure}

\begin{table}
\caption{Mass accreted by the first-formed WDs in DD progenitor systems in our sample that experience a combination of RLOF and wind mass transfer (1366 for the standard case; 1213 for the optimistic case), split by donor type. The bottom row gives the SSS life-time of the average $\Delta M$, if all material is accreted at the component-specific (for H and He, respectively) steady-burning mass-transfer rates, and the resulting expected number of accreting systems (with Poisson errors). Both columns use the $\gamma$-CE prescription.}
 \centering
  \begin{tabular}{c c c c}
  \hline
					& standard		& upper limit	\\
					& SeBa case:		& case:	\\
  donor stellar				& $\Delta M$/SN		& $\Delta M$/SN		\\
  type					& [M$_{\odot}$]		& [M$_{\odot}$]		\\
  \hline
  \hline
   main sequence			& 0.0 			& 1.04 $\cdot 10^{-4}$ \\
   Herzsprung gap			& 0.0 			& 2.74 $\cdot 10^{-5}$ \\
   first giant branch			& 1.20 $\cdot10^{-4}$ 	& 1.86 $\cdot 10^{-5}$ \\
   core He-burning			& 0.0 			& 3.12 $\cdot 10^{-4}$ \\
   asymptotic giant branch		& 0.0 			& 6.63 $\cdot 10^{-5}$ \\
   He-star				& 8.35 $\cdot10^{-5}$ 	& 3.26 $\cdot 10^{-3}$ \\
   He-giant				& 1.43 $\cdot10^{-1}$ 	& 2.01 $\cdot 10^{-1}$ \\
   total, all types			& 1.43 $\cdot10^{-1}$	& 2.05 $\cdot 10^{-1}$ \\
  \hline
   $\tau_{\mathrm{accr}}$ [yr]		& 5.8 $\cdot10^{4}$	& 8.42 $\cdot10^{4}$	\\
   $N_{\mathrm{accr}}$  ($10^{10}$ L$_{\mathrm{B,}\odot}$ galaxy)	& 1.7 $\cdot10^{2} \pm$ 13	& 2.5 $\cdot10^{2} \pm$ 16	\\
  \hline
\end{tabular} \label{table:deltaM_donor_types_RLOF}
\end{table}
Figure \ref{fig:accr_history_log_4_RLOF} shows the accretion history of these systems.

Table \ref{table:time_from_last_accr_to_SN_-_RLOF} gives the time from the last mass transfer event until the SN explosion for all the systems experiencing mass transfer via RLOF in our sample, for the optimistic assumptions. For systems transferring mass via RLOF, $\sim$1\% of the systems explode less than 10 Myrs after the cessation of mass transfer.

\begin{table}
\caption{Time from last mass transfer event to SN explosion, systems experiencing mass transfer via RLOF at least once, optimistic case. Compare with tables \ref{table:time_from_last_accr_to_SN_-_all} and \ref{table:time_from_last_accr_to_SN_-_wind}.}
 \centering
  \begin{tabular}{c c c}
  \hline
  time since			& no. of	& fraction of		\\
  last accr			& systems	& total			\\
  \hline
   t $<$ 1 Myr			& 2		& 1.65 $\cdot10^{-3}$	\\
   1 Myr $< t <$ 10 Myr		& 14		& 1.15 $\cdot10^{-2}$	\\
   10 Myr $< t <$ 100 Myr	& 61		& 5.03 $\cdot10^{-2}$	\\
   100 Myr $< t <$ 400 Myr	& 77		& 6.35 $\cdot10^{-2}$	\\
   t $>$ 400 Myr		& 1059		& 8.73 $\cdot10^{-1}$	\\
  \hline
  Total				& 1213		& 1.00			\\  
  \hline
\end{tabular} \label{table:time_from_last_accr_to_SN_-_RLOF}
\end{table}

%

\section{Discussion} \label{sect:Discussion}
The aim of this study was to determine whether DD progenitor systems of type Ia SNe could conceivably constitute a significant population of SSSs during the SD-like part of their evolution. If they would, then the observationally inferred absence of SSSs would strongly limit the DD progenitor scenario. Our results indicate that DD progenitors do not make up a significant population of SSSs for either of the cases we have examined. As mentioned in Section \ref{sect:Method}, the mass transfer and retention efficiencies assumed in the optimistic case are probably not realistic, but even with such optimistic assumptions, we estimate a total galactic population of just 6-7 wind accreting proto-DD SN progenitors in large spiral galaxies like the Milky Way. In contrast, the study by Di Stefano (\cite{Di.Stefano.2010.b}) predicted 'thousands' of wind-accreting SSS proto-DD progenitors. The number of systems undergoing RLOF that could be SSSs under our optimistic assumptions is larger, around $\sim$ 250, but still quite negligible compared to Di Stefano's estimate. For the standard SeBa case, we find no wind accreting SSSs, and $\sim$ 170 SSSs from RLOFing progenitors. We stress that all numbers estimated in the optimistic case should be considered generous upper limits. The cause of the disagreement between the results of our study and that of our comparison study may be the somewhat general assumptions concerning the donor mass loss rates used by Di Stefano, although this can hardly explain the difference completely. As mentioned, that study predicts thousands) of donor stars capable of supplying a large mass loss rate, and combined with a large enough retention efficiency this leads to an appreciable population of SSS proto-DD type Ia SN progenitor systems accreting over a long period of time. But it is unclear to the authors of this article how such a large population of long-lived, large $\dot{M}$-donors arises, and a similar population is not reproduced by SeBa, despite using the same accretion efficiency.

Our study predicts a different chemical composition of the accreted material for proto-DD accretors. Contrary to what is expected for the SD scenario, helium mass transfer dominates the SD-like phase of proto-DD type Ia supernova progenitors. As mentioned, this has a significant effect on the maximal life-times of any SSS phases of such systems. This is not just a feature of the optimistic case; also for the standard case, the vast majority of the accreted material is helium.

In our study, the average masses accreted in systems experiencing RLOF ($\sim$0.2 M$_{\odot}$) is similar to the accreted masses hypothesised for the most massive carbon-oxygen WDs in the SD scenario. However, the fact that the majority of the accreted material is helium results in significantly shorter SSS life-times. Our results therefore hinge on the details concerning the mass transfer and steady-burning of helium, which are currently less well-understood than for hydrogen. However, due to the higher temperatures and densities required for helium-burning, the steady-burning rate that we have assumed is probably not unreasonable.

The studies by Di Stefano (\cite{Di.Stefano.2010.a}) and Gilfanov \& Bogdan (\cite{Gilfanov.Bogdan.2010}) indicated that the observed numbers of SSSs in nearby galaxies and the integrated supersoft luminosity in ellipticals are one to two orders of magnitude too small compared to what should be expected if luminous SSS SD progenitors were the main contributors to the type Ia SN rate. According to our study, we expect a factor 10-20 fewer SSS DD progenitors compared to SSS SD progenitors, for the same SN rate. If we accept the more constraining case, i.e. that the difference between the observed number and/or integrated luminosities of SSSs fall two orders of magnitude short of the expected value, then the lack of SSSs would also be a problem for the DD progenitor scenario, at least under the assumption that all accreted material is burned at the steady-burning rate, although it would still be a lot smaller problem than for the SD scenario. However, one should be careful, as the long delay between the SSS phase and the actual explosion would make a direct comparison between the current SSS population and the current type Ia SN rate impossible. What would then be needed is to take the details of the star formation history of the galaxies for which the SSS populations have been determined into account and use these to calculate the expected SSS population for the DD scenario. That is beyond the scope of the present paper.

Because of the time required for the second WD to form in a DD progenitor system, we expected to find the cessation of accretion long before the merger that leads to the type Ia SN. The accretion history plots in Figures \ref{fig:accr_history_log_4_all}-\ref{fig:accr_history_log_4_RLOF} and Tables \ref{table:time_from_last_accr_to_SN_-_all}-\ref{table:time_from_last_accr_to_SN_-_RLOF} show that even if our optimistic assumptions were correct, the SSS behaviour would have ceased millions of years before the SN explosion for most systems. We note that systems accreting exclusively via a wind generally stop accreting earlier (with respect to the merger) than systems accreting via a combination of wind and RLOF.

The applicability of our results depends on the correctness of the assumptions on which SeBa is based. For the evolution towards type Ia SNe, Toonen et al. (\cite{Toonen.et.al.2012}) and Bours et al. (\cite{Bours.et.al.2013}) found that the most crucial assumptions for the SD channel are the retention efficiency and the CE-prescription, whereas the DD channel is relatively insensitive to the latter assumption. Our results explicitly vary the retention efficiency (from realistic to extremely optimistic values in the two cases) and we note that our results hardly depend on which CE-formalism is used.

The SN rate inferred from earlier observations is approximately a factor 10 larger than what can be produced with the current version of SeBa with DD progenitors (Toonen et al. \cite{Toonen.et.al.2012}), although recent observations indicate that the observed rate may be smaller than previously thought, and therefore the discrepancy between the simulated DD populations may fall correspondingly less short, to within a factor of a few of the observed type Ia SN rate (Bours et al. \cite{Bours.et.al.2013}). This means that, theoretically, our results could underestimate the numbers of SSS progenitors for the DD scenario. However, if the DD scenario really is the dominant progenitor scenario the discrepancy between the theoretical rates and the observed rates must be either due to an incorrect normalisation with the correct binary evolution channels, or to completely new binary evolution channels that have yet to be identified. Although difficult to prove, we believe the discrepancy is more likely due to a normalisation issue, and certainly there is no strong reason to believe that any additional DD binary evolution channels would produce very long SSS phases.  Whatever the exact reason for and magnitude of the discrepancy, in our analysis we implicitly assumed that we can scale our results from the current (too small) SeBa DD type Ia rate to the actual SN rate, to compare with SSS SD progenitors.

On a fundamental note, we emphasise that the discussion concerning SSS usually implicitly assumes that such sources are more or less 'naked', i.e. unobscured by local material. If, for whatever reason, the sources in question are significantly obscured by local matter (as would likely be the case if the systems were transferring mass via a wind, where a large fraction of the matter lost from the donor would not be accreted onto the accretor), the situation may well be different (see Nielsen et al. \cite{Nielsen.et.al.2013b}). That local, circumbinary material may have been present around at least some type Ia SNe immediately prior to the explosion has been established by a number of studies (Gerardy et al. \cite{Gerardy.et.al.2004}, Borkowski et al. \cite{Borkowski.et.al.2006}, Patat et al. \cite{Patat.et.al.2007}, Sternberg et al. \cite{Sternberg.et.al.2011}, Chiotellis et al. \cite{Chiotellis.et.al.2011}).

%

\section{Conclusions} \label{sect:Conclusion}
The key to solving the type Ia SN progenitor question is a better understanding of the observational characteristics of the accretion process which eventually brings the accreting WD to the mass required for the SN. While observational data of the SNe themselves is rapidly growing as a result of several large-scale SN surveys, the theoretical understanding of what a nuclear burning WD looks like remains a sticky point. Without a better grasp on this issue, the riddle of the SN Ia progenitors is going to remain unsolved, presumably until such a time when direct confirmation of a (very) nearby SN Ia progenitor can be made.

We performed an analysis of the accretion behaviour of DD type Ia SN progenitor systems in the evolutionary stage prior to the formation of the second WD, where the systems may conceivably be similar to SD type Ia SN progenitors, and hence possibly SSSs. For this, we simulated 500,000 binary systems with the population synthesis code SeBa. We made our analysis for two cases: 1) a conservative / realistic case using the standard SeBa assumptions concerning mass-transfer and retention efficiencies, and 2) a more optimistic case using less constrained assumptions for wind accretion (i.e. 25\% mass-transfer efficiency, 100\% retention efficiency) to establish firm upper limits on the possible SSS behaviour of proto-DD type Ia SN progenitor systems. For both cases, we calculated the average accreted mass per SN, the corresponding SSS life-time, and the expected number of accreting SSSs in a Milky-Way type galactic population at any given time, assuming all mass transfer happens at the rate required for steady burning for the type of material in question (hydrogen or helium).

For wind accretion we observe the following:
   \begin{enumerate}
      \item In the standard case, no mass is accreted via wind, so we expect no SSS behaviour at all for DD progenitors of type Ia SNe if the standard SeBa assumptions concerning wind accretion are generally correct.
      \item For the relaxed assumptions in the optimistic case, the average mass accreted per supernova via pure wind mass transfer is tiny.
      \item Unlike what is likely the case for SD progenitors, the majority of the accreted material is helium, not hydrogen. Even if this material were accreted at the steady-burning rate (which it most likely is not) it would not be sufficient to make the systems luminous SSSs for very long; in our estimate, the average SSS lifetime is of the order of a couple of thousands of years. Translated into numbers of systems in a Milky-Way type galaxy, this corresponds to less than 10 luminous SSSs originating from proto-DD type Ia SN progenitors accreting via a wind.
   \end{enumerate}

The conclusions we make from this is that we do not expect a significant number of proto-DD type Ia SN progenitor to be observable as SSSs as a result of pure wind mass transfer. This goes contrary to what was concluded in the study by Di Stefano (\cite{Di.Stefano.2010.b}), which predicted thousands of these sources..\\

For systems transferring mass via a wind and RLOF, the basic conclusions are rather similar to those concerning pure wind-accreting systems, with a few modifications:
    \begin{enumerate}
     \item For both the standard and optimistic cases, the average masses accreted per supernova may be significantly (more than an order of magnitude) larger than for pure wind-accreting systems.
     \item As for wind accreting systems, the vast majority of the transferred mass is helium.
     \item The length of the supersoft X-ray emitting phase for these systems will be of the order of $10^{5}-10^{4}$ years, if all mass transfer happens at the steady-burning rates. While this is significantly longer than for pure wind-accreting systems, it is still negligible compared to the total life-time of DD type Ia SN progenitor systems, and at least an order of magnitude smaller than the expected SSS life-time of SD progenitors. The expected number of accreting systems present in a galactic population is correspondingly smaller.
    \end{enumerate}

To sum up: on the basis of our study, we do not find support for the existence of a significant population of SSS proto-DD type Ia SN progenitor systems. This holds for both wind- and RLOF-accreting systems, although the tendency is stronger for wind accretors. Since no SSSs are expected if the DD progenitor scenario of type Ia SNe, the absence of observed SSSs is not a strong argument against the DD progenitor scenario.

%

\newpage
\clearpage

\begin{acknowledgements}

This research is supported by NWO Vidi grant 016.093.305. We thank the referee for a careful reading, and many constructive suggestions.

\end{acknowledgements}

\begin{appendix}
\clearpage

\section{Appendix}

\begin{table}
 \caption{Mass accreted by the first-formed WDs in all DD progenitor systems (3298 for the standard case; 2920 for the optimistic case), split by donor type. The bottom row gives the SSS life-time of the average $\Delta M$, if all material is accreted at the component-specific (for H and He, respectively) steady-burning mass-transfer rates. Both columns use the $\alpha$-CE prescription.}
 \centering
  \begin{tabular}{c c c}
  \hline
					& standard		& upper limit		\\
					& SeBa case:		& case:			\\
  donor stellar				& $\Delta M$/SN		& $\Delta M$/SN		\\
  type					& [M$_{\odot}$]		& [M$_{\odot}$]		\\
  \hline
  \hline				
   main sequence star			& 0.0				& 4.61 $\cdot10^{-5}$ \\ 
   Herzsprung gap star			& 0.0				& 3.64 $\cdot10^{-5}$ \\ 
   first giant branch star		& 5.77 $\cdot10^{-5}$		& 4.39 $\cdot10^{-5}$ \\ 
   core He-burning star			& 0.0				& 8.05 $\cdot10^{-5}$ \\ 
   asymptotic giant branch star	& 0.0				& 1.62 $\cdot10^{-5}$ \\ 
   He-star				& 8.27 $\cdot10^{-4}$		& 3.10 $\cdot10^{-3}$ \\ 
   He-giant star			& 8.46 $\cdot10^{-2}$		& 1.09 $\cdot10^{-1}$ \\ 
   Total, all types			& 8.54 $\cdot10^{-2}$		& 1.13 $\cdot10^{-1}$ \\
  \hline
  \hline
   $\tau_{\mathrm{accr}}$ [yr]		& 3.4 $\cdot10^{4}$		& 4.6 $\cdot10^{4}$	\\ 
   $N_{\mathrm{accr}}$ ($10^{10}$ L$_{\mathrm{B,}\odot}$ galaxy)	& 1.0 $\cdot10^{2}$	& 1.4 $\cdot10^{2}$	\\
  \hline
\end{tabular} \label{table:deltaM_donor_types_all_alphaCE}
\end{table}

\begin{table}
\caption{Mass accreted via wind mass transfer by the first-formed WDs in DD progenitor systems in our sample (1384 for the standard case; 1363 for the optimistic case), split by donor type. The bottom row gives the SSS life-time of the average $\Delta M$, if all material is accreted at the component-specific (for H and He, respectively) steady-burning mass-transfer rates. Both columns use the $\alpha$-CE prescription.}
 \centering
  \begin{tabular}{c c c}
  \hline
					& standard		& upper limit	\\
					& SeBa case:		& case:	\\
  donor stellar				& $\Delta M$/SN		& $\Delta M$/SN		\\
  type					& [M$_{\odot}$]		& [M$_{\odot}$]		\\
  \hline
  \hline
   main sequence star			& 0.0			& 0.0 			\\ 
   Herzsprung gap star			& 0.0			& 3.51 $\cdot10^{-5}$ 	\\ 
   first giant branch star		& 0.0			& 4.37 $\cdot10^{-5}$ 	\\ 
   core He-burning star			& 0.0			& 2.28 $\cdot10^{-6}$ 	\\ 
   asymptotic giant branch star	& 0.0			& 1.86 $\cdot10^{-6}$ 	\\ 
   He-star				& 0.0			& 9.84 $\cdot10^{-4}$ 	\\ 
   He-giant star			& 0.0			& 3.00 $\cdot10^{-3}$ 	\\ 
   total, all types			& 0.0			& 4.07 $\cdot10^{-3}$	\\
  \hline
  \hline
   $\tau_{\mathrm{accr}}$ [yr]		& 0.0			& 2.0 $\cdot 10^{3}$	\\ 
   $N_{\mathrm{accr}}$ ($10^{10}$ L$_{\mathrm{B,}\odot}$ galaxy)	& none	& 6.0	\\
  \hline
\end{tabular} \label{table:deltaM_donor_types_wind_alphaCE}
\end{table}

\begin{table}
\caption{Mass accreted by the first-formed WDs in DD progenitor systems in our sample that experience a combination of RLOF and wind mass transfer (1914 for the standard case; 1557 for the optimistic case), split by donor type. The bottom row gives the SSS life-time of the average $\Delta M$, if all material is accreted at the component-specific (for H and He, respectively) steady-burning mass-transfer rates. Both columns use the $\alpha$-CE prescription.}
 \centering
  \begin{tabular}{c c c}
  \hline
					& standard		& upper limit	\\
					& SeBa case:		& case:	\\
  donor stellar				& $\Delta M$/SN		& $\Delta M$/SN		\\
  type					& [M$_{\odot}$]		& [M$_{\odot}$]		\\
  \hline
  \hline
   main sequence star			& 0.0			& 8.64 $\cdot 10^{-5}$	\\ 
   Herzsprung gap star			& 0.0			& 3.75 $\cdot 10^{-5}$	\\ 
   first giant branch star		& 9.95 $\cdot 10^{-5}$	& 4.40 $\cdot 10^{-5}$	\\ 
   core He-burning star			& 0.0			& 1.49 $\cdot 10^{-4}$	\\ 
   asymptotic giant branch star	& 0.0			& 2.87 $\cdot 10^{-5}$	\\ 
   He-star				& 1.42 $\cdot 10^{-3}$	& 4.96 $\cdot 10^{-3}$	\\ 
   He-giant star			& 1.46 $\cdot 10^{-1}$	& 2.02 $\cdot 10^{-1}$	\\ 
   Total, all types			& 1.48 $\cdot 10^{-1}$	& 2.08 $\cdot 10^{-1}$	\\
  \hline
  \hline
   $\tau_{\mathrm{accr}}$ [yr]		& 5.9 $\cdot 10^{4}$	& 8.5 $\cdot 10^{4}$	\\ 
   $N_{\mathrm{accr}}$ ($10^{10}$ L$_{\mathrm{B,}\odot}$ galaxy)	& 1.8 $\cdot 10^{2}$	& 2.5 $\cdot 10^{2}$	\\
  \hline
\end{tabular} \label{table:deltaM_donor_types_RLOF_alphaCE}
\end{table}

\newpage
\clearpage

\end{appendix}


\begin{thebibliography}{}

  \bibitem[1983]{Abt.1983}
      Abt, H.~A., 1983,
      ARA\&A, 21, 343

  \bibitem[2006]{Borkowski.et.al.2006}
      Borkowski, K.~J., Hendrick, S.~P. \& Reynolds, S.~P., 2006
      ApJ, 652, 1259

  \bibitem[2013]{Bours.et.al.2013}
      Bours, M.~C.~P., Toonen, S. \& Nelemans, G.
      A\&A, accepted

  \bibitem[2011]{Chiotellis.et.al.2011}
      Chiotellis, A., Schure, K.~M., \& Vink, J., 2011,
      A\&A, 537, A139

  \bibitem[2004]{Edgar.2004}
      Edgar, R., 2004,
      New Astronomy Reviews, 48, 843

  \bibitem[2004]{Gerardy.et.al.2004}
      Gerardy, C.~L., Hoeflich, P., Fesen, R.~A., Marion, G.~H., Nomoto, K., Quimby, R., Schaefer, B.~E., Wang, L. \& Wheeler, C., 2004
      ApJ, 607, 391

  \bibitem[1979]{Fujimoto.Sugimoto.1979}
      Fujimoto, M.~Y. \& Sugimoto, D., 1979,
      PASJ, 31, 1

  \bibitem[1982]{Fujimoto.Sugimoto.1982}
      Fujimoto, M.~Y. \& Sugimoto, D., 1982,
      ApJ, 257, 291

  \bibitem[2010]{Gilfanov.Bogdan.2010}
      Gilfanov, M. \& Bogd{\'a}n, {\'A}., 2010,
      Nature, 463, 924

  \bibitem[1999]{Hachisu.et.al.1999}
      Hachisu, I., Kato, M., Nomoto, K., \& Umeda, H., 1999,
      ApJ, 519, 314

  \bibitem[1975]{Heggie.1975}
      Heggie, D.~C., 1975,
      MNRAS, 173, 729

  \bibitem[1992]{van.den.Heuvel.et.al.1992}
      van den Heuvel, E.~P.~J., Bhattacharya, D., Nomoto, K., \& Rappaport, S.~A., 1992,
      A\&A, 262, 97

  \bibitem[1984]{Iben.Tutukov.1984}
      Iben, Jr., I. \& Tutukov, A.~V., 1984,
      ApJS, 54, 335

  \bibitem[1996]{Iben.Tutukov.1996}
      Iben, Jr., I. \& Tutukov, A.~V., 1996,
      ApJs, 105, 145

  \bibitem[2013]{Ivanova.et.al.2013}
      Ivanova, N., Justham, S., Chen, X., De Marco, O., Fryer, C.~L., Gaburov, E., Ge, H., Glebbeek, E., Han, Z., Li, X.-D., Lu, G., Marsh, T., Podsiadlowski, P., Potter, A., Soker, N., Taam, R., Tauris, T.~M., van den Heuvel, E.~P.~J. \& Webbink, R.~F., 2013,
      A\&ARv, 21, 59

  \bibitem[2011]{Kashi.Soker.2011}
      Kashi, A. \& Soker, N., 2011, MNRAS, 417, 1466

  \bibitem[1999]{Kato.Hachisu.1999}
      Kato, M. \& Hachisu, I., 1999,
      ApJL, 513, L41

  \bibitem[1987]{de.Kool.et.al.1987}
      de Kool, M., van den Heuvel, E.~P.~J. \& Pylyser, E., 1987,
      A\&A, 183, 47

  \bibitem[1993]{Kroupa.et.al.1993}
      Kroupa, P., Tout, C.~A. \& Gilmore, G., 1993,
      MNRAS, 262, 545

  \bibitem[2012]{Maoz.Mannucci.2012}
      Maoz, D. \& Mannucci, F., 2012,
      PASA, 29, 2012

  \bibitem[2010]{Maoz.et.al.2010}
      Maoz, D., Sharon, K., \& Gal-Yam, A., 2010,
      ApJ, 722, 1879

  \bibitem[2007]{Mohamed.Podsiadlowski.2007}
      Mohamed, S. \& Podsiadlowski, P., 2007,
      ASP-CS, 372, 397

  \bibitem[2011]{Mohamed.Podsiadlowski.2011}
      Mohamed, S. \& Podsiadlowski, P., 2011,
      ASP-CS, 445, 355

  \bibitem[2000]{Nelemans.et.al.2000}
      Nelemans, G., Verbunt, F., Yungelson, L.~R. \& Portegies Zwart, S.~F., 2000,
      A\&A, 360, 1011

  \bibitem[2001]{Nelemans.et.al.2001}
      Nelemans, G., Yungelson, L.~R., Portegies Zwart, S.~F. \& Verbunt, F., 2001,
      A\&A, 365, 491

  \bibitem[2012]{Nielsen.et.al.2012}
      Nielsen, M., Voss, R. \& Nelemans, G., 2012,
      MNRAS, 426, 2668

  \bibitem[accepted]{Nielsen.et.al.2013a}
      Nielsen, M., Voss, R. \& Nelemans, G.
      MNRAS, accepted, doi: 10.1093/mnras/stt1250

  \bibitem[2013]{Nielsen.et.al.2013b}
      Nielsen, M.~T.~B., Dominik, C., Nelemans, G. \& Voss, R., 2013
      A\&A, 549, A32

  \bibitem[1979]{Nomoto.et.al.1979}
      Nomoto, K., Nariai, K. \& Sugimoto, D. 1979,
      PASJ, 31, 287-298

  \bibitem[1982]{Nomoto.1982}
      Nomoto, K., 1982,
      ApJ, 253, 798

  \bibitem[2007]{Patat.et.al.2007}
      Patat, F., Chandra, P., Chevalier, R., Justham, S., Podsiadlowski, P., Wolf, C., Gal-Yam, A., Pasquini, L., Crawford, I.~A., Mazzali, P.~A., Pauldrach, A.~W.~A., Nomoto, K., Benetti, S., Cappellaro, E., Elias-Rosa, N., Hillebrandt, W., Leonard, D.~C., Pastorello, A., Renzini, A., Sabbadin, F., Simon, J.~D. \& Turatto, M., 2007,
      Science, 317, 924

  \bibitem[1993]{Phillips.1993}
      Phillips, M.~M. 1993,
      ApJL, 413, L105

  \bibitem[1996]{Portegies.Zwart.1996}
      Portegies Zwart, S.~F. \& Verbunt, F., 1996,
      A\&A, 309, 179

  \bibitem[2010a]{Di.Stefano.2010.a}
      Di Stefano, R., 2010,
      ApJ, 712, 728

  \bibitem[2010b]{Di.Stefano.2010.b}
      Di Stefano, R., 2010,
      ApJ, 719, 474

  \bibitem[2011]{Sternberg.et.al.2011}
      Sternberg, A., Gal-Yam, A., Simon, J.~D., Leonard, D.~C., Quimby, R.~M., Phillips, M.~M., Morrell, N., Thompson, I.~B., Ivans, I., Marshall, J.~L., Filippenko, A.~V., Marcy, G.~W., Bloom, J.~S., Patat, F., Foley, R.~J., Yong, D., Penprase, B.~E., Beeler, D.~J., Prieto, C.~A. and Stringfellow, G.~S., 2011
      Science, 333, 856

  \bibitem[2012]{Toonen.et.al.2012}
      Toonen, S., Nelemans, G. \& Portegies Zwart, S., 2012,
      A\&A, 546, A70

  \bibitem[in prep.]{Toonen.et.al.2013}
      Toonen, S., Nelemans, G., Bours, M., S. Portegies Zwart \& Claeys, J., 2013,
      arXiv e-prints 1302.0495

  \bibitem[1979]{Tutukov.Yungelson.1979}
      Tutukov, A. \& Yungelson, L., 1979,
      A\&A, 83, 401

  \bibitem[2008]{Voss.Nelemans.2008}
      Voss, R. \& Nelemans, G., 2008
      Nature, 451, 802

  \bibitem[1984]{Webbink.1984}
      Webbink, R.~F., 1984,
      ApJ, 277, 355

  \bibitem[1973]{Whelan.Iben.1973}
      Whelan, J. \& Iben, Jr., I. 1973,
      ApJ, 186, 1007-1014

\end{thebibliography}
\end{document}